\documentclass[letterpaper]{article}

\usepackage[T1]{fontenc}

\usepackage{geometry}
\usepackage{setspace}

\usepackage[style = chem-acs]{biblatex}
\usepackage{hyperref}
\usepackage{graphicx}
\usepackage{float}
\usepackage{subcaption}
\newfloat{scheme}{htbp}{los}
\floatname{scheme}{Scheme}
\floatname{chart}{Chart}
\newfloat{graph}{htbp}{loh}

\usepackage{chemformula} 
\usepackage[version = 4]{mhchem} 

\usepackage{authblk}
\author[1]{Abhishek Das}
\author[2, 3]{Neelesh Kumar Vij}
\author[1, 4]{Demitry Farfurnik}
\affil[1]{Department of Electrical and Computer Engineering, NC State University, Raleigh, NC-27606, USA}
\affil[2]{Institute of Research in Electronics and Applied Physics and Joint Quantum Institute, University of Maryland, College Park, Maryland 20740, United States}
\affil[3]{Department of Electrical and Computer Engineering, University of Maryland, College Park, Maryland 20740, United State}
\affil[4]{Department of Physics and Astronomy, NC State University, Raleigh, NC-27606, United State}

\title{Optimization of circular cavities via guided-mode expansion method based inverse design}
\date{*Email: adas24@ncsu.edu}

\begin{document}

\maketitle

\begin{abstract}
  Spin-photon interfaces, realized by coupling optically active spin systems to photonic cavities, are essential for quantum networking and quantum information processing. Implementing such an interface for polarization-encoded photons requires a cavity that supports arbitrary polarization, provides efficient optical access through its far-field mode, and maintains sufficiently high quality factors to enable high cooperativity with the system's optical transitions. However, inherent trade-offs between the Q-factor and far-field emission mode make the simultaneous optimization of these parameters toward the realization of spin-photon interfaces challenging. In this work, we implement a gradient-based inverse-design framework using guided-mode expansion with automatic differentiation to obtain the geometrical features of a circular ring cavity that supports arbitrary polarization while simultaneously optimizing the cavity quality factor and far-field mode profile. The resulting optimized non-periodic cavity achieves a quality factor of approximately $9,000$, about an order-of-magnitude higher than that of a periodic ("bullseye") cavity while preserving a Gaussian-like far-field emission pattern. Furthermore, by varying the cavity geometry within a $\pm 6$ nm fabrication tolerance, we demonstrate the robustness of the design against fabrication errors and identify the innermost ring width and central disk radius as the parameters with the greatest impact on the quality factor and far-field mode. These results establish guided mode expansion-based inverse design as a powerful and computationally efficient approach for developing high-cooperativity spin-photon interfaces for quantum photonic applications.
\end{abstract}

\section*{Keywords}

Spin-photon interface, photonic cavities, quantum photonics, inverse design



\section{Introduction}

Spin-photon interfaces, which coherently couple an optically active matter qubit acting as a deterministic single-photon emitter to a photonic cavity, are foundational building blocks in quantum networking and distributed quantum computing applications. \cite{2023NatRP...5..326C, 2016NaPho..10..631A}  In such applications, quantum information encoded in a stationary matter qubit must be faithfully transferred to a flying photonic qubit for transmission over optical channels, enabling protocols such as entanglement distribution between distant nodes, \cite{knaut2024entanglement, PRXQuantum.5.010202}  quantum key distribution, \cite{PRXQuantum.5.010202, basso2021qkd} and generation of multi-dimensional cluster states for measurement-based quantum computing. \cite{cogan2023cluster, thomas2022graph} 

The efficient implementation of a spin-photon interface relies on a cavity that satisfies several key requirements. First, to enable the deterministic generation of entanglement between polarization-encoded photons, the cavity must support arbitrary photon polarization, \cite{Gudat2011PermanentTuning, Bakker2011PolarizationDegenerate} which can be achieved by utilizing cavity geometries with full rotational symmetry. \cite{Bauch2024TailoredCavity, Mehdi2024PolarisationRotations, Wijitpatima2024CircularBragg} Second, for the high-fidelity transfer of quantum information from the matter qubit to a photonic qubit, the cavity must exhibit a sufficiently high quality factor (Q-factor) that leads to a high emitter-cavity cooperativity.  \cite{Thompson1992ObservationNormal, Hood1998RealTime, HJKimble_1998, Yoshie2004} Third, to efficiently collect photons emitted by the system into a well-defined free space or fiber mode, the cavity must exhibit an efficient far-field emission profile. \cite{PhysRevLett.111.237403, press2008control, barnes2002singlephoton, Zwiller2002Solid} However, a fundamental trade-off exists between the requirements on the Q-factor and far-field emission mode: efficient out-coupling of photons requires the cavity mode to leak into the light cone, which in turn decreases the Q-factor. Consequently, cavities with high Q-factors that may achieve high emitter-cavity cooperativity typically exhibit highly divergent far-field emission patterns that are poorly matched to the collection optics. \cite{wei2012completedeterministicmultielectrongreenbergerhornezeilinger, sun2018SinglePhot} Meanwhile, cavities that support arbitrary polarization and offer an efficient, nearly Gaussian far-field emission pattern such as the periodic circular Bragg cavity (also known as the “bullseye cavity”), typically exhibit Q-factors only up to the order of $\sim 1000$. \cite{Davanco2011Bullseye, Zheng:17, Singh2022OpticalTransparency} Consequently, engineering a cavity that supports arbitrary polarization and simultaneously achieves a high Q-factor to obtain high emitter-cavity cooperativity and an accessible, directional far-field emission mode, represents a central challenge for scalable quantum networking hardware.

In this work, we design a circular symmetric cavity of concentric rings (for supporting arbitrary polarization) that satisfies the competing objectives of achieving a high Q-factor and a directional far-field. Since the exhaustive parameter sweep over all the geometrical features is highly impractical, we obtain the optimal cavity geometry by utilizing an inverse design algorithm, in which an automated optimization drives the cavity geometry directly toward a targeted set of performance metrics. \cite{Molesky_2018} Among the available inverse design strategies, including genetic algorithms, \cite{minkov2014optimization} neural network-based approaches,\cite{AsanoNoda+2019+2243+2256} and gradient-based methods, \cite{Minkov2020NanoLett, vij2025inversedesignedphotoniccrystalcavities} we choose a gradient-based optimizer well suited to problems with large number of design degrees of freedom, as the gradient computation cost is independent of the number of parameters. \cite{Piggott_2015} Since the explicit calculation of the electromagnetic fields in the cavity via finite-difference-time-domain (FDTD) simulations is time and resource-consuming, our gradient-based inverse design algorithm employs the guided-mode expansion (GME) method that solves Maxwell's equations by expanding the electromagnetic field in terms of the guided modes of an effective slab, \cite{PhysRevB.73.235114, Minkov2020NanoLett} consistent with previous works optimizing cavity Q-factor and far-field modes that did not consider the support of arbitrary polarization. GME is a frequency-domain method, hence the Q-factor is directly accessible as an eigenvalue imaginary part, thereby providing a substantial computational speedup over the time-domain analysis required in FDTD. \cite{Minkov2020NanoLett} However, since GME is an approximate method, our validation of the performance of the final cavity design involves additional FDTD simulations. 

While the GME-based inverse-design approach is applicable to any material platform, we simulate the performance of our optimized cavity on a GaAs substrate. We choose GaAs as a host of InAs quantum dots that can feature a spin qubit, emit pure and indistinguishable single photons, and due to the straightforward fabrication processes in GaAs, can support cavity-emitter coupling with high cooperativity. \cite{2007NaPho...1..215S, PhysRevLett.86.1502, warburton2013spins, Singh2022OpticalTransparency} 

The optimized non-periodic cavity obtained by our inverse design framework achieves a Q-factor of $\sim 9,000$, nearly an order-of-magnitude improvement over the one of the periodic bullseye cavity in the same material platform, \cite{Singh2022OpticalTransparency, Davanco2011Bullseye, Zheng:17} while preserving a highly directional far-field emission pattern that has an $87\%$ mode overlap with a Gaussian beam that represents a collection lens with a standard NA of 0.68. By further varying the geometrical features of the inverse-designed cavity by $\pm 6 $nm, we show that the obtained cavity is robust against fabrication errors,  while identifying the innermost ring width and central disk radius as the most fabrication-sensitive features. Our results demonstrate that GME-based inverse design provides a computationally efficient and broadly applicable pathway toward high-cooperativity spin-photon interfaces, and we anticipate that combining this framework with state-of-the-art quantum dot fabrication will bring within reach the cooperativity levels required for scalable quantum networking applications.

\begin{figure}[htbp]
    \centering

    \begin{subfigure}[b]{0.28\textwidth}
        \centering
        \includegraphics[width=\textwidth]{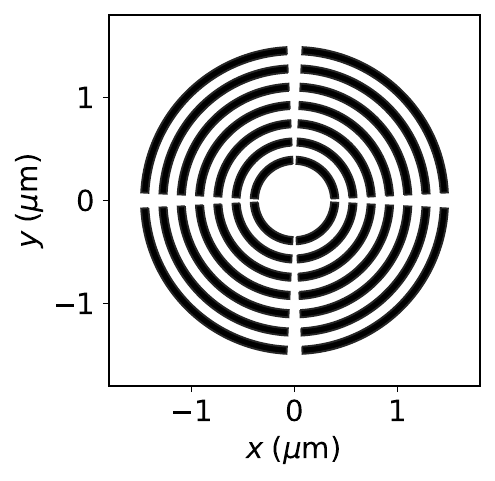}
        \caption{}
        \label{Original Bullseye}
    \end{subfigure}\hfill
        \begin{subfigure}[b]{0.35\textwidth}
        \centering
        \includegraphics[width=\textwidth]{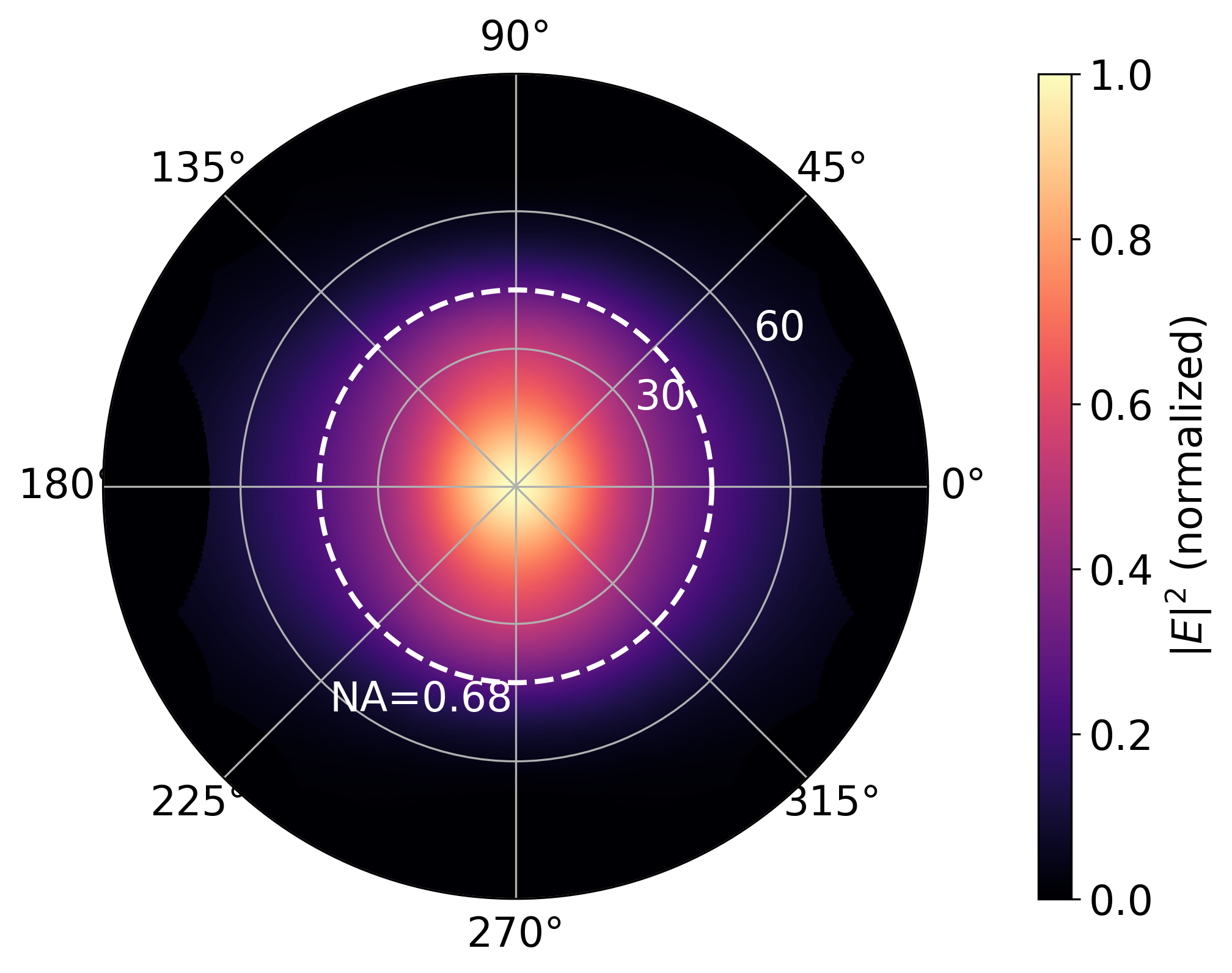}
        \caption{}
        \label{Far_Field_Original_Bullseye_GME}
    \end{subfigure}
    \begin{subfigure}[b]{0.35\textwidth}
        \centering
        \includegraphics[width=\textwidth]{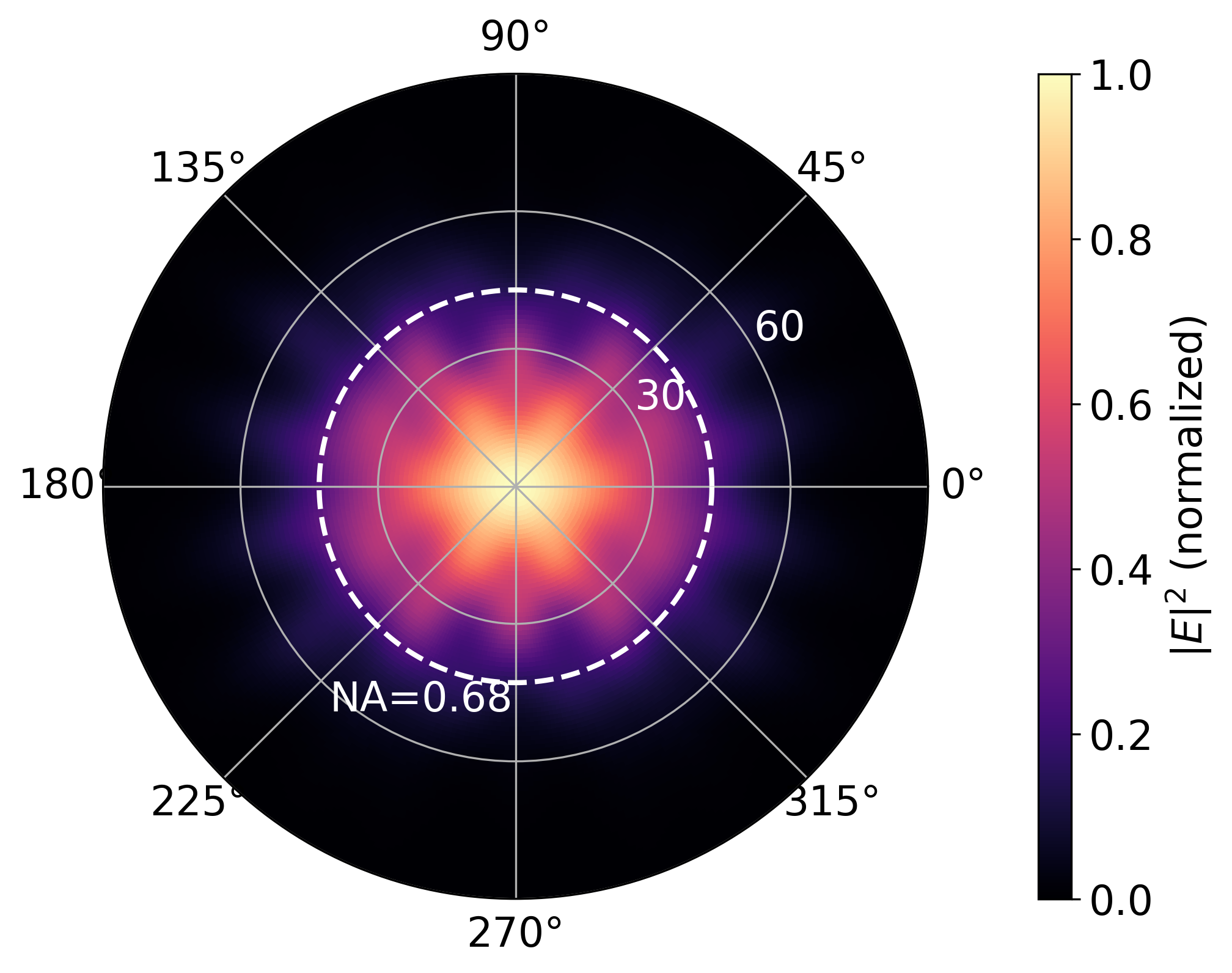}
        \caption{}
        \label{Far_Field_Original_Bullseye_FDTD}
    \end{subfigure}\hfill

    \caption{(a) Illustration of the periodic bullseye cavity, top (x-y) view. The black areas illustrate the etched portion, the rest is GaAs substrate. (b)-(c) The far-field emission mode of the periodic bullseye cavity, computed in (b) GME by calculating a discrete set of radiation channels via reciprocal lattice vectors and interpolating the result to a continuous plot and (c) FDTD by a near-to-far-field transformation.}
    \label{Original Bullseye figures}
\end{figure}

\section{Results and discussion}

Before using the GME method for the inverse design of cavities, we begin by validating the method against explicit FDTD simulations of the standard, periodic bullseye cavity with the structural dimensions fabricated in ref. \cite{Singh2022OpticalTransparency} on GaAs. Figure \ref{Original Bullseye} shows the cross-sectional (x-y plane) view of the periodic bullseye that consists of concentric rings, with narrow bridges dividing each ring into four quadrants. These bridges provide structural support to prevent the cavity from collapsing (as a result of undercutting the sacrificial layer below the cavity during fabrication) and facilitate electrical contact that allows the charging of the quantum dots with single electrons or holes for spin-photon interfacing.  \cite{LiNvCenter}

To simulate the performance of the periodic bullseye cavity, we utilize the Legume package \cite{ZANOTTI2024109286} that provides GME-based analysis of photonic crystal slabs with automatic differentiation capabilities. In Legume, we model the periodic bullseye cavity as a GaAs slab (permittivity $\epsilon = 12.25$, thickness $d = 0.20a$) embedded in a square periodic supercell of side $L = 8a$, where $a$ is the length unit. The cavity consists of a central disk of radius $r_0 = 0.355a$ and N = 7 concentric air-filled rings ($\epsilon=1$). Each ring has a radial width, $w_1 = ... =w_7 = 0.07a$, and consecutive rings are separated by a pitch of $d_1 = ... = d_6 = 0.1075a$. \cite{Davanco2011Bullseye, Ates_2012, Singh2022OpticalTransparency, Sapienza_2015} The supercell side $L$ is chosen to be significantly larger than the grating footprint ($ \sim 1.4a$) so that the artificial periodicity has a negligible effect on the mode frequencies and the radiation patterns. We set the unit length $a$ to be $1 \mu m$ for concreteness, however, since Maxwell's equations are scale invariant, all computed quantities rescale trivially to any other choice of $a$. Considering the cavity geometry, we use both the GME approach using Legume and FDTD method using the software Tidy3D to simulate two parameters. The first parameter is the $Q$-factor of the cavity and the second parameter is the overlap of the far-field mode of the cavity with a Gaussian profile that matches an objective lens of numerical aperture (NA) of $0.68$, which is a typical objective lens used in experimental setups that involve such cavities. \cite{Singh2022OpticalTransparency} The GME simulation of the periodic bullseye cavity results in a Q-factor of $1,364$, in a good quantitative agreement with corresponding FDTD result of $1,212$. Furthermore, both simulations result in a nearly Gaussian far-field mode (Figures \ref{Far_Field_Original_Bullseye_GME} and \ref{Far_Field_Original_Bullseye_FDTD}) with mode overlaps of $94\%$ and $96\%$, respectively. Despite a small deviation of the FDTD result (Figure \ref{Far_Field_Original_Bullseye_FDTD}) from a smooth Gaussian profile that arises from the high resolution (40 mesh points per wavelength) of the simulation, the agreement between the far-field modes achieved via GME and FDTD confirms the validity of the GME method, and motivates us to use the method for cavity inverse design.

Assured of the reliability of our GME simulation method, we now leverage the method to implement the gradient descent based inverse-design algorithm for the optimization of the circular cavity. Starting with the  dimensions of the periodic bullseye cavity shown in  Figure \ref{Original Bullseye figures}, the algorithm iteratively modifies eight geometrical parameters of the cavity, the radius of the central disk, $r_0$, the radial widths of the individual rings $w_1, w_2, ..., w_7$ to achieve simultaneously high Q-factor and far-field mode overlap. While ensuring that the resonance frequency is still within the range of emission of InAs/GaAs quantum dots of $\sim 930$ nm, the GME method iteratively calculates the gradients of the geometrical parameters to minimize a loss function, which is a function of weighted sum of three terms:
\begin{equation}
    \mathcal{L} = 1- w_Q\cdot \tau_Q - w_{FF} \cdot \tau_{FF} - w_f \cdot \tau_f .
    \tag{1} 
    \label{loss-equation}
\end{equation}
Here, $\tau_f, \tau_{FF}, \tau_Q$ are terms associated with the frequency, far-field mode overlap, and Q-factor with relative weights of $w_f = 0.20, w_{FF}, w_Q = 0.55, w_{FF} = 0.25$, respectively.
Mathematically, the frequency term is defined as a Lorentzian, 
\begin{equation}
    \tau_f = \frac{1}{1+ \left(\frac{f-f_{target}}{\Delta f} \right)^2},
    \tag{2}
    \label{frequency-term}
\end{equation}
where $f_{target} = 1.09(c/a)$ and $\Delta f$ is the half-width of the Lorentzian, $0.03(c/a)$, $c$ is the speed of light, and $a$ is the length unit in the GME simulation. The specific values of $f_{target}$ and $\Delta f$ correspond to the typical InAs/GaAs quantum dot emission wavelength in the 910-980 nm regime and the smooth form of the Lorentzian ensures a persistent gradient throughout. 
Meanwhile, the far-field mode overlap term in the loss function is defined as the squared amplitude overlap between the normalized simulated far-field mode and a normalized Gaussian reference profile, 
\begin{equation}
    \tau_{FF} = \left \lvert \sum_j {E}_{sim} \left(\vec{k}_j \right) \cdot {E}_{Gauss} \left(\vec{k}_j \right)\right \rvert^2,
    \tag{3}
    \label{Far-field term}
\end{equation}
where the sum runs over all the radiation channels within the light cone, and $E_{Gauss}$ is the reference Gaussian profile, $E_{Gauss} \propto \exp \left[-\left(\frac{\sin \theta}{NA}^2 \right) \right] \cdot \sqrt{\cos \theta}$. The term $\sqrt{\cos \theta}$ is the apodization factor, which arises from energy conservation in the Richards-Wolf vectorial diffraction framework \cite{RichardsApodization}.
Finally, the term associated with the Q-factor in the loss function is given by
\begin{equation}
    \tau_Q = A \cdot \min \left[\frac{\log \left(1 + \frac{Q}{Q_{ref}} \right)}{\log 2}, 1 \right].
    \tag{4}
    \label{Q-term}
\end{equation}
Here, the logarithmic form compresses the contribution the Q-factor into a bounded, smooth function that saturates to unity at a predefined value, $Q_{ref}$, and $A$ is a numerical factor representing the overlap of the eigenvector of the calculated cavity mode with the mode of the initial cavity, which ensures that the optimizer does not converge on spurious modes with large, unphysical Q-factors arising from the finite reciprocal lattice truncation in GME. \cite{PhysRevB.73.235114, PhysRevB.83.085301}

\begin{figure}[htbp]
    \centering
    \begin{subfigure}{0.9\textwidth}
        \centering
        \includegraphics[width = \textwidth]{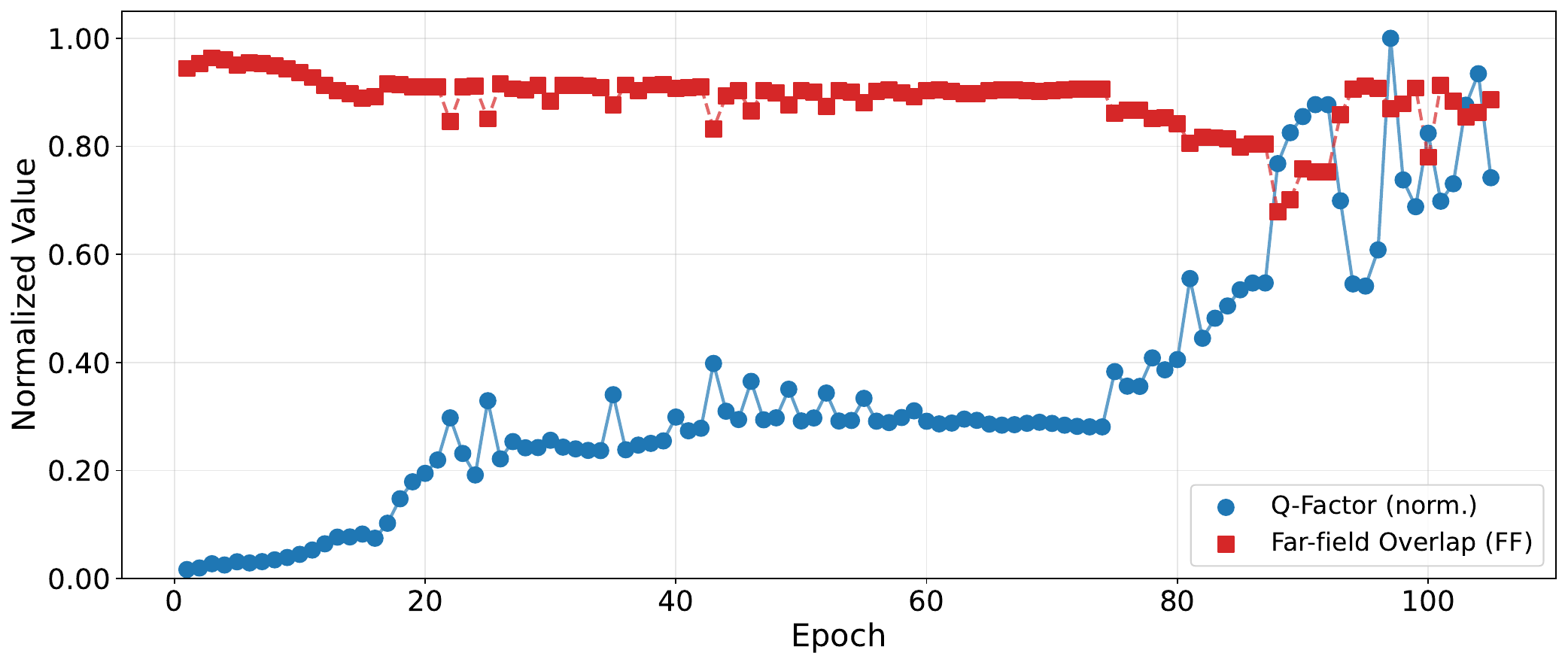}
        \caption{}
        \label{evolution-over-epochs}
    \end{subfigure}
    \begin{subfigure}{0.32\textwidth}
        \centering
        \includegraphics[width=\textwidth]{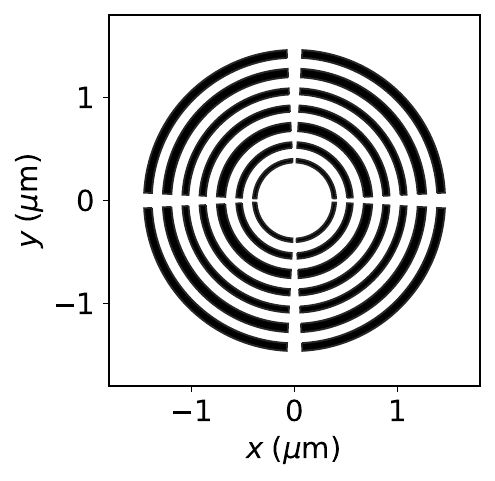}
        \caption{}
        \label{Inverse_Designed_Bullseye_Geometry}
    \end{subfigure}
    \begin{subfigure}{0.38\textwidth}
        \centering
        \includegraphics[width=\textwidth]{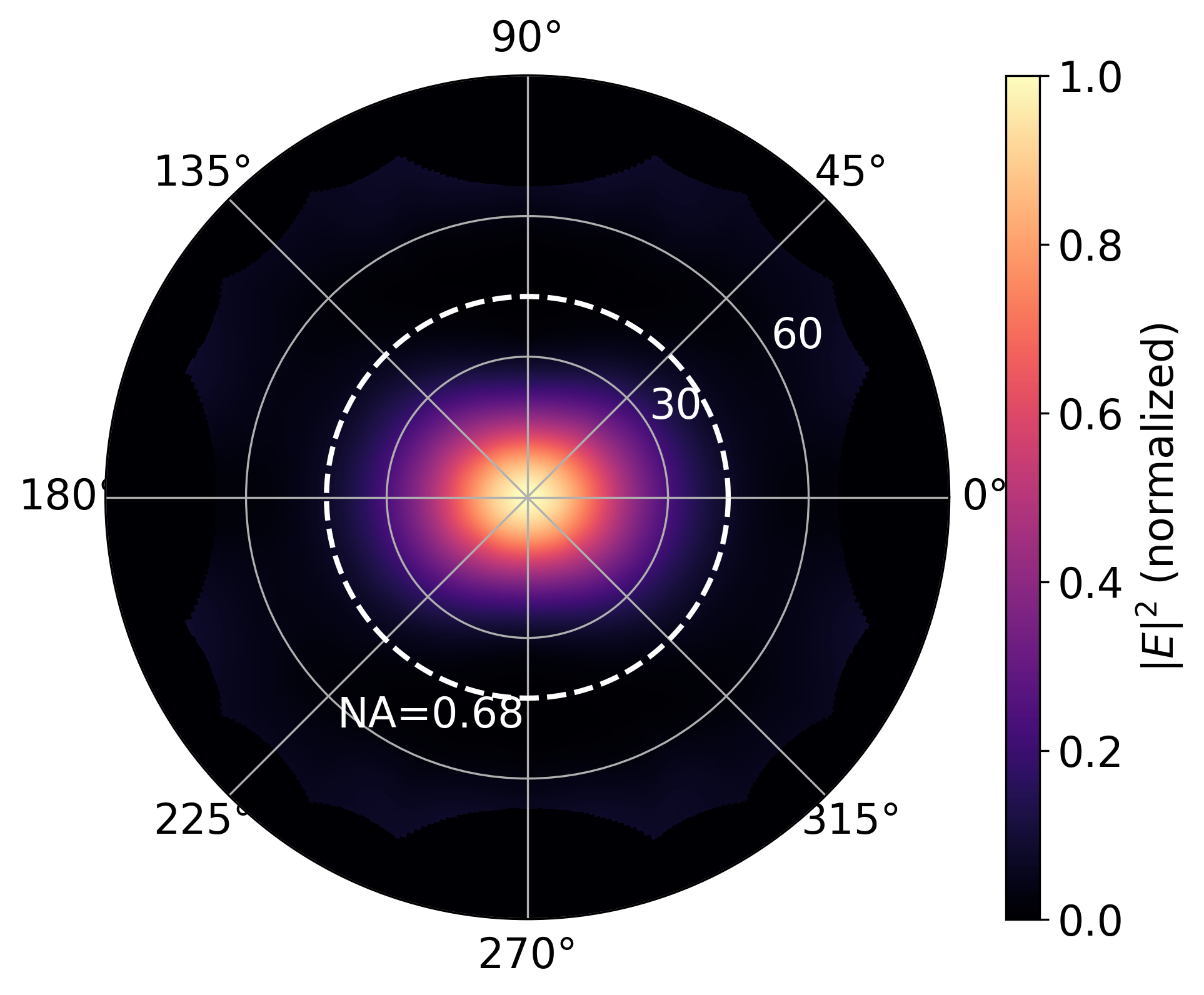}
        \caption{}
        \label{Inverse_Designed_Bullseye_Far_Field_GME}
    \end{subfigure}
    \caption{Optimization of the circular cavity geometry via inverse design. (a) The key metrics of gradient descent inverse design for the optimization of the circular cavity as a function of the number of epochs. The blue circles represent the evolution of the Q-factor (Normalized to the highest value of all the epochs) and the red squares represent the far-field mode overlap with a Gaussian function. The Q -factor increases steadily and reaches its maximum at around $\sim 105$ epochs, while the far-field mode overlap remains consistently high throughout the epochs. (b) View (x-y plane) of the optimized cavity geometry, black region representing the etched areas in the GaAs slab. (c) The far-field emission pattern of the inverse-designed cavity simulated via GME and normalized to the peak intensity .}
    \label{Inverse-Design-GME}
\end{figure}

After designing the framework, we run the gradient descent algorithm toward the optimization of the geometry of the circular cavity for a total of 105 epochs. Figure \ref{evolution-over-epochs} illustrates the evolution of the far-field mode overlap and Q-factor across all epochs. The full optimization was structured in seven stages with gradually increasing values of the Q-factor target parameter, $Q_{ref}$, since starting the simulation with an overly ambitious target Q-factor does not lead to practical improvements in the calculation of its gradient. As illustrated by the red squares in Figure \ref{evolution-over-epochs}, the far-field mode overlap remains high $(> 0.7)$ throughout the epochs in all seven stages. Meanwhile, the Q-factor (blue circles in Figure \ref{evolution-over-epochs}) gradually increases until reaching saturation at epoch 105. At the end of the optimization process, the GME algorithm results in a novel non-periodic cavity of dimensions $r_0 = 369 \, \mu$m and $\left(w_1, w_2, w_3, w_4, w_5, w_6, w_7 \right) =$  $(34$ nm, $55$ nm, $81$ nm, $61$ nm, $59$ nm, $82$ nm, $75$ nm) (Figure \ref{Inverse_Designed_Bullseye_Geometry}) that substantially improves the Q-factor while maintaining a nearly Gaussian far-field emission mode (Figure \ref{Inverse_Designed_Bullseye_Far_Field_GME}). 

\begin{figure}[htbp]
    \centering

    \begin{subfigure}[b]{0.40\textwidth}
        \centering
        \includegraphics[width=\textwidth]{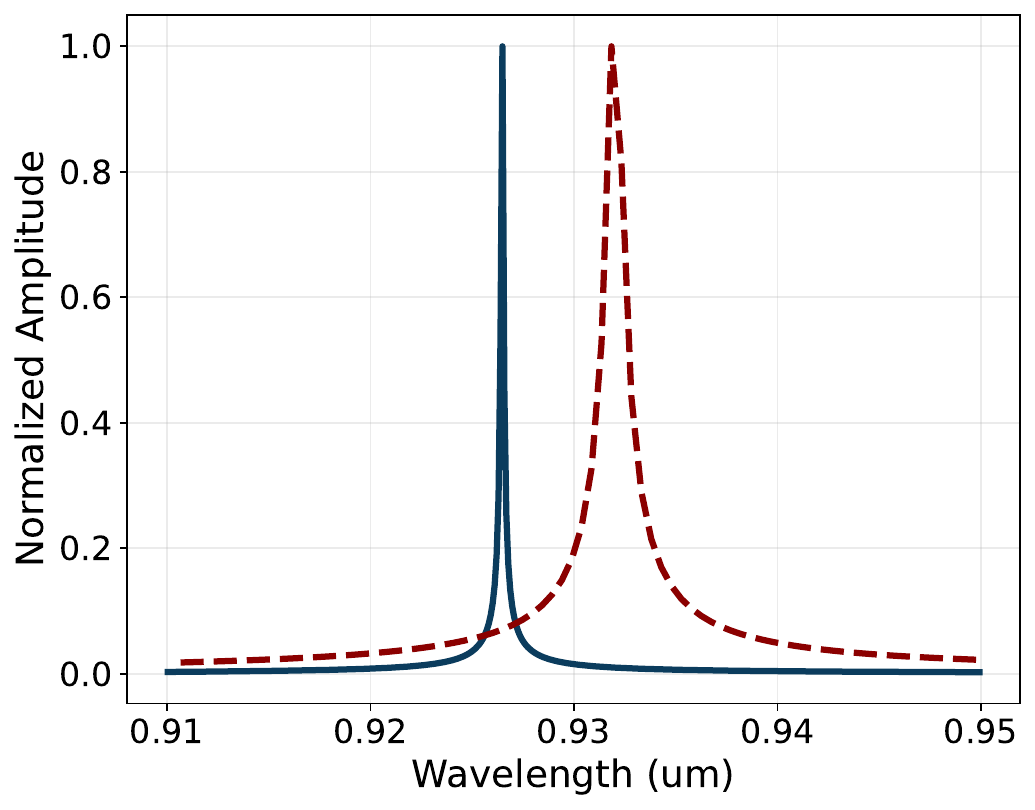}
        \caption{}
        \label{Spectrum-comparison}
    \end{subfigure}
        \begin{subfigure}[b]{0.40\textwidth}
        \centering
        \includegraphics[width=\textwidth]{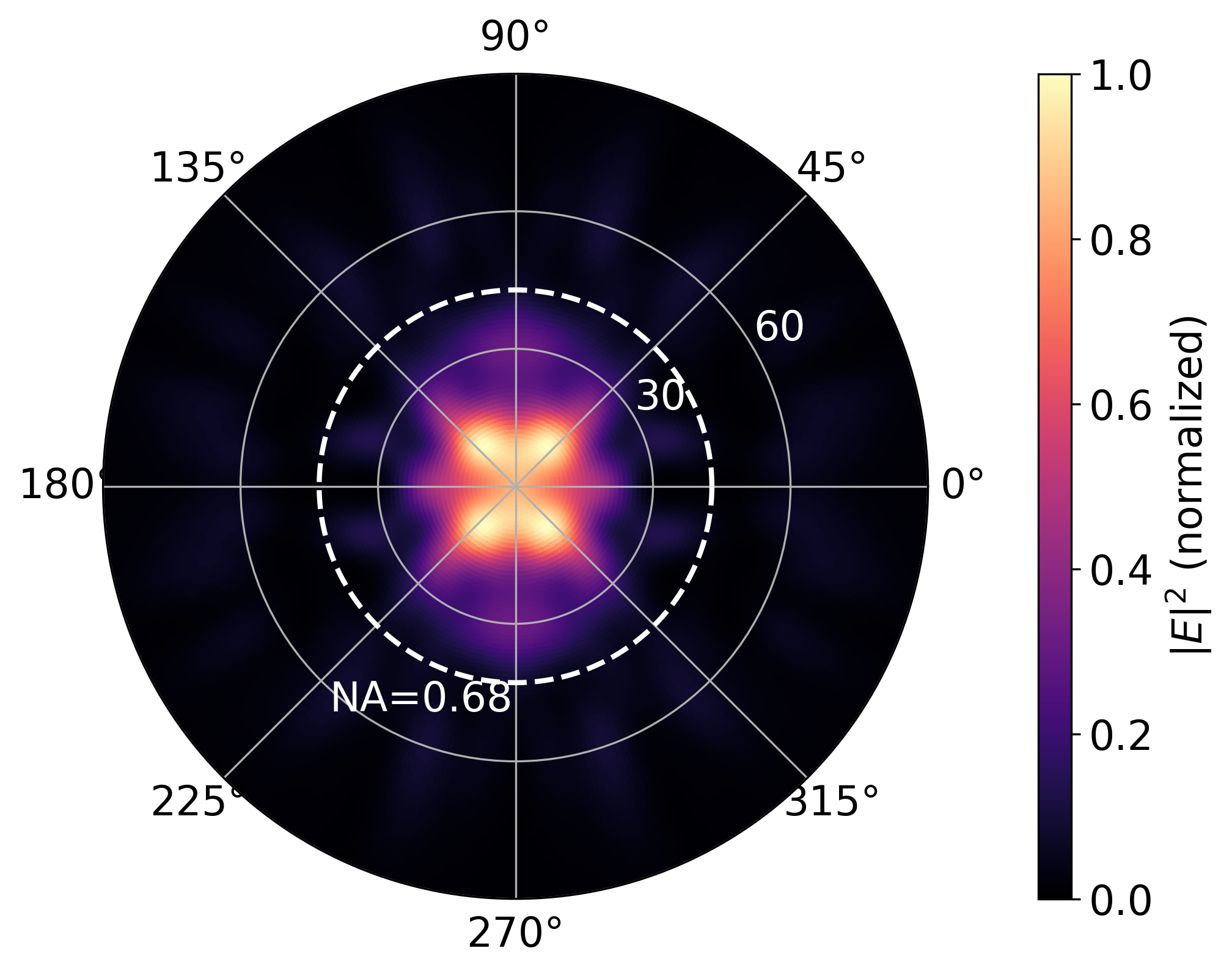}
        \caption{}
        \label{inverse-designed-far-field}
    \end{subfigure}
    \begin{subfigure}[b]{0.40\textwidth}
        \centering
        \includegraphics[width=\textwidth]{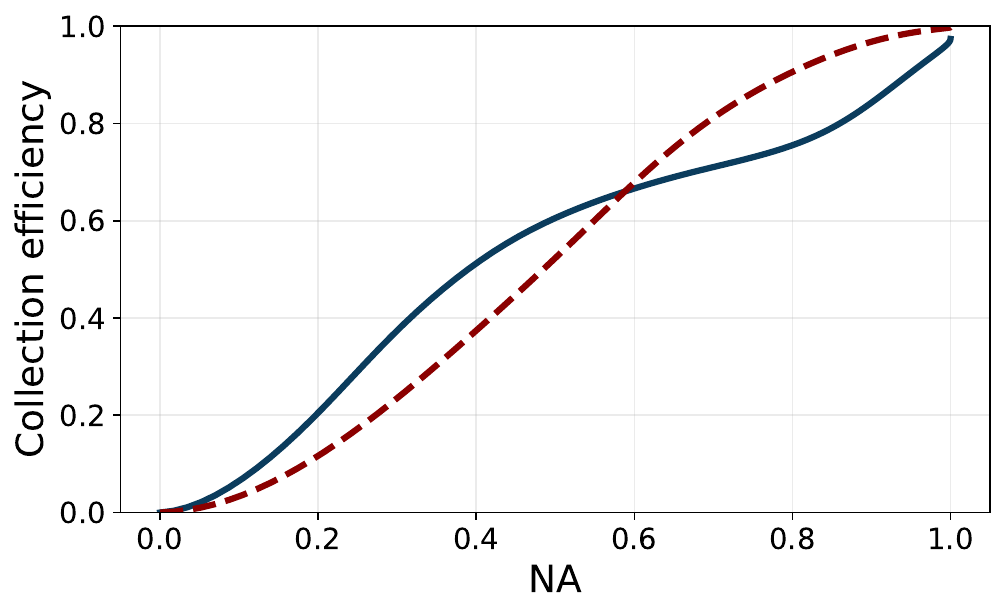}
        \caption{}
        \label{Collection-Efficiency}
    \end{subfigure}\hfill

    \caption{FDTD simulations of the inverse-designed cavity. (a) The normalized resonance spectra of the inverse-designed cavity (solid blue, $Q \sim 9,000$) and the periodic bullseye cavity (dashed red, $Q \sim 1,200$), showing the narrower linewidth achieved by cavity optimization. (b) The far-field emission pattern of the inverse-designed cavity normalized to the peak intensity. (c) Collection efficiency as a function of NA for the inverse-designed (solid blue) and periodic bullseye (dashed red) cavities at their respective resonance wavelengths.}
    \label{Inverse-Designed-figures-FDTD}
\end{figure}

Since Q-factor results obtained by GME are only an approximation, \cite{PhysRevB.73.235114, ZANOTTI2024109286} we validate the inverse-designed cavity by simulating its performance using FDTD (Figure \ref{Inverse-Designed-figures-FDTD}). \cite{Das2026BullseyeTidy3D} First, the FDTD simulation of the spectral response of the cavity (solid blue line in Figure \ref{Spectrum-comparison}) exhibits a sharp response at $\sim 930$ nm with a Q-factor of $9,000$, almost an order-of-magnitude larger than the Q-factor of $\sim 1,200$ of the periodic bullseye cavity (dashed red line in Figure \ref{Spectrum-comparison}), with a slight wavelength shift due to the geometric modifications introduced by the optimizer. Second, the FDTD simulation of the far-field emission mode of the inverse-designed cavity (illustrated in Figure \ref{inverse-designed-far-field}) exhibits a nearly Gaussian profile, with small deviations from the GME simulation (Figure \ref{Inverse_Designed_Bullseye_Far_Field_GME}) and a smooth Gaussian profile arising from the high resolution (40 mesh points per wavelength) of the simulation. Quantitatively, the far-field mode overlaps with an ideal Gaussian beam collected by an objective lens of NA = 0.68 by $87 \%$, slightly lower than the $96 \%$ overlap of a periodic bullseye cavity (Figure \ref{Far_Field_Original_Bullseye_FDTD}).

To further compare the performance of the inverse-designed cavity with the periodic bullseye cavity, we calculate the optical collection efficiencies provided by both cavities. We define the collection efficiency as the ratio between the optical power collected by a lens above the cavity and the total optical power emitted from the cavity upward. Figure \ref{Collection-Efficiency} compares the simulated collection efficiency as a function of the NA of the lens for both cavities at their respective resonance wavelengths. These simulations show that the collection efficiency provided by the inverse-designed cavity (solid blue line in Figure \ref{Collection-Efficiency}) is greater than the one provided by the periodic bullseye cavity (dashed red line in Figure \ref{Collection-Efficiency}) across the entire low-to-moderate NA range. The two curves cross near NA = 0.6, above which the periodic bullseye cavity provides a slightly greater collection efficiency (consistent with the slightly higher mode overlap of the far-field mode with a Gaussian) before both converge toward unity at NA = 1.0. The superior performance of the inverse-designed cavity at NA values below 0.6 is particularly relevant for applications involving low-NA collection channels such as coupling into lensed fibers, \cite{tran2025efficientfibercouplingtelecom, Langer2025} where maximizing the fraction of power emitted into a narrow cone is critical. Overall, the collection efficiency simulations indicate that the multi-objective loss function used in our inverse design protocol guided the resulting cavity toward a far-field mode where light is collected efficiently, while simultaneously achieving an order-of-magnitude improvement in the Q-factor over the periodic cavity.

\begin{figure}[htbp]
    \centering
    \begin{subfigure}{0.45\textwidth}
        \centering
        \includegraphics[width=\textwidth]{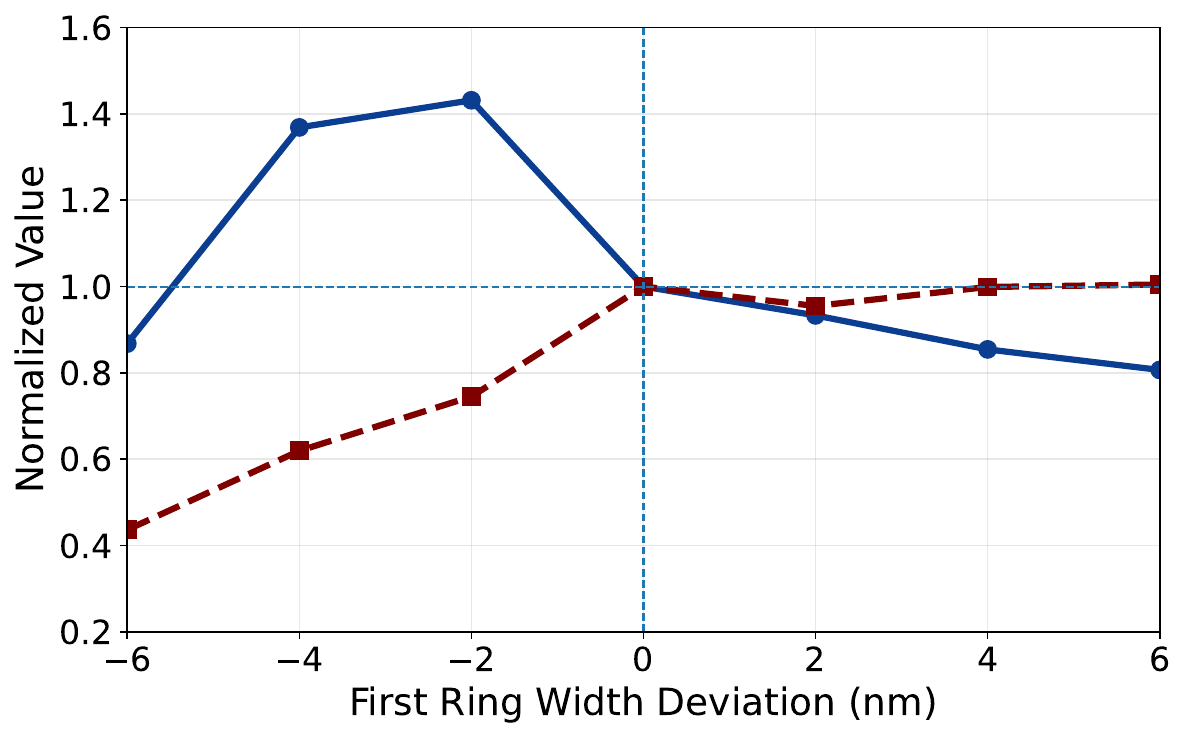}
        \caption{}
        \label{first-ring-width}
    \end{subfigure}\hfill
    \begin{subfigure}[b]{0.45\textwidth}
        \centering
        \includegraphics[width=\textwidth]
        {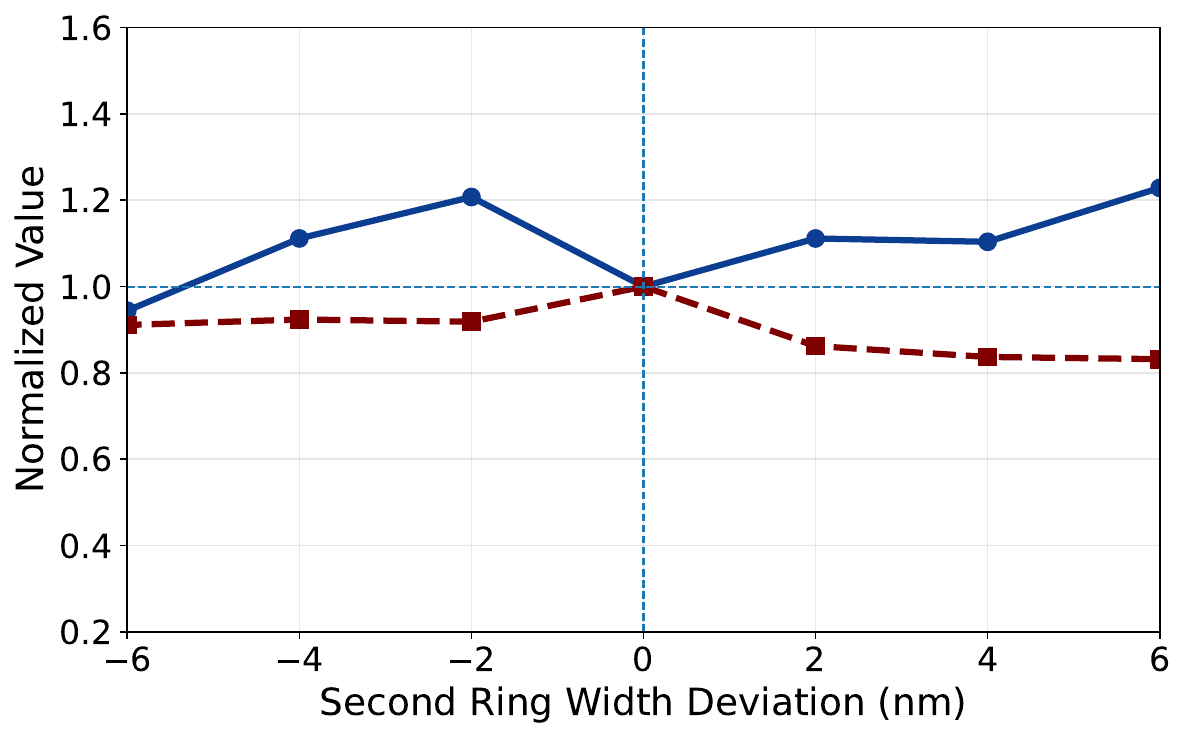}
        \caption{}
        \label{second-ring-width}
    \end{subfigure}
    \vspace{0.8 cm}
    \begin{subfigure}{0.45\textwidth}
        \centering
        \includegraphics[width=\textwidth]{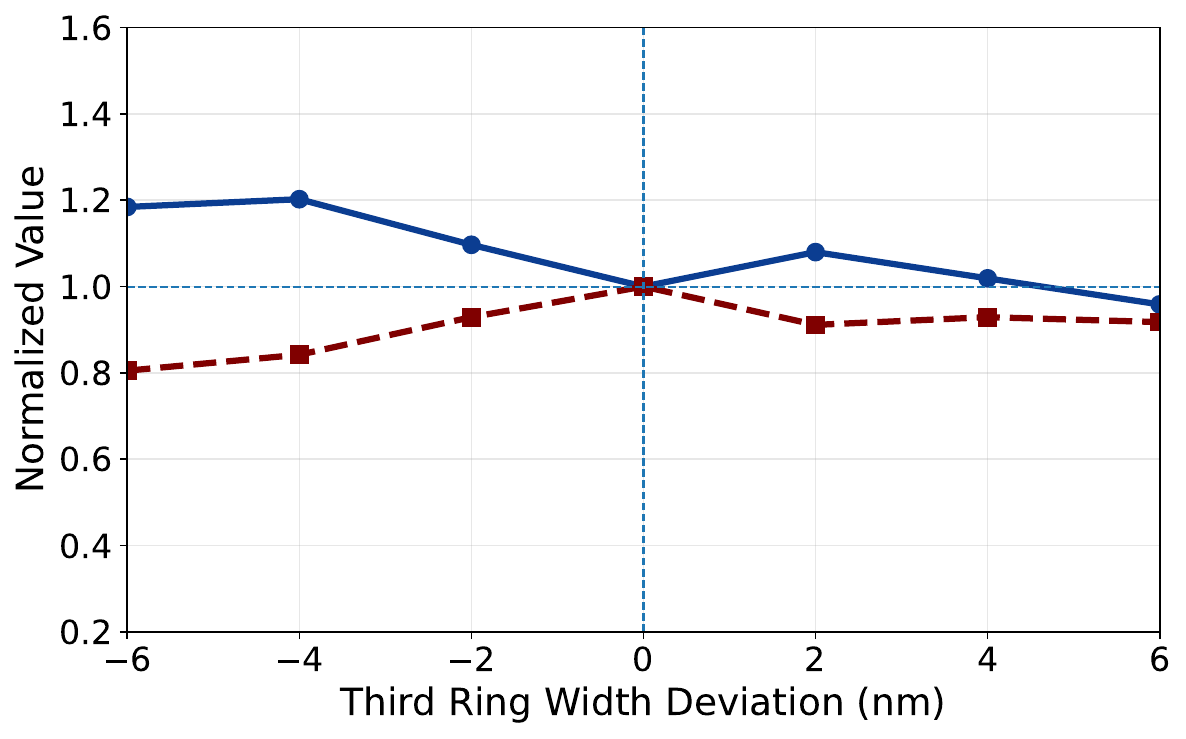}
        \caption{}
        \label{third-ring-width}
    \end{subfigure}\hfill
    \begin{subfigure}[b]{0.45\textwidth}
        \centering
        \includegraphics[width = \textwidth]{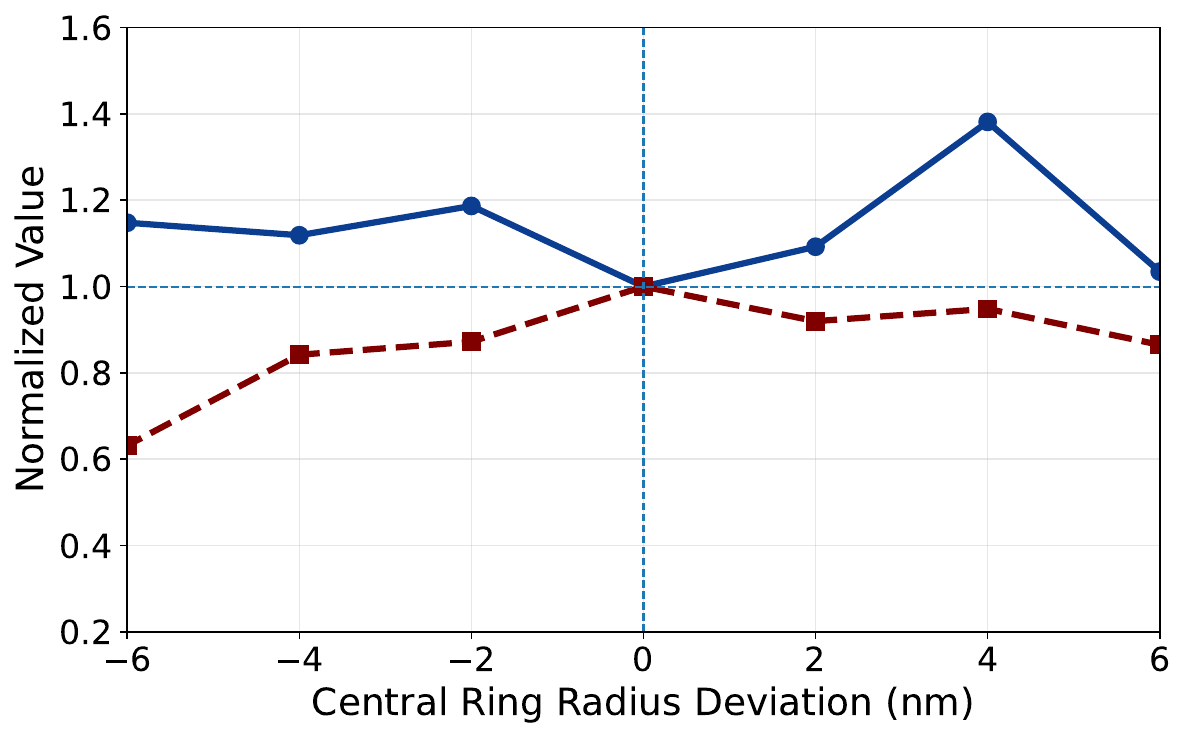}
        \caption{}
        \label{central-ring-radius}
    \end{subfigure}
    \vspace{0.8 cm}
    \begin{subfigure}{0.45\textwidth}
        \centering
        \includegraphics[width=\textwidth]{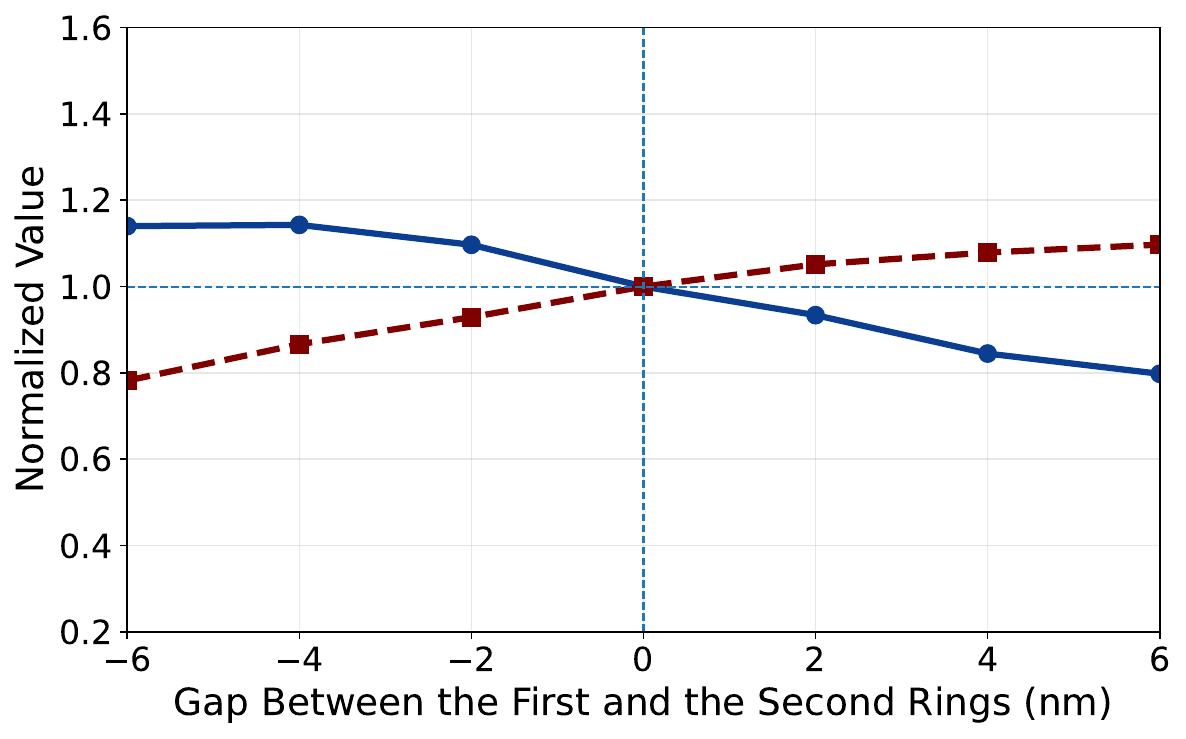}
        \caption{}
        \label{first-ring-gap}
    \end{subfigure}\hfill
    \begin{subfigure}[b]{0.45\textwidth}
        \centering
        \includegraphics[width=\textwidth]
        {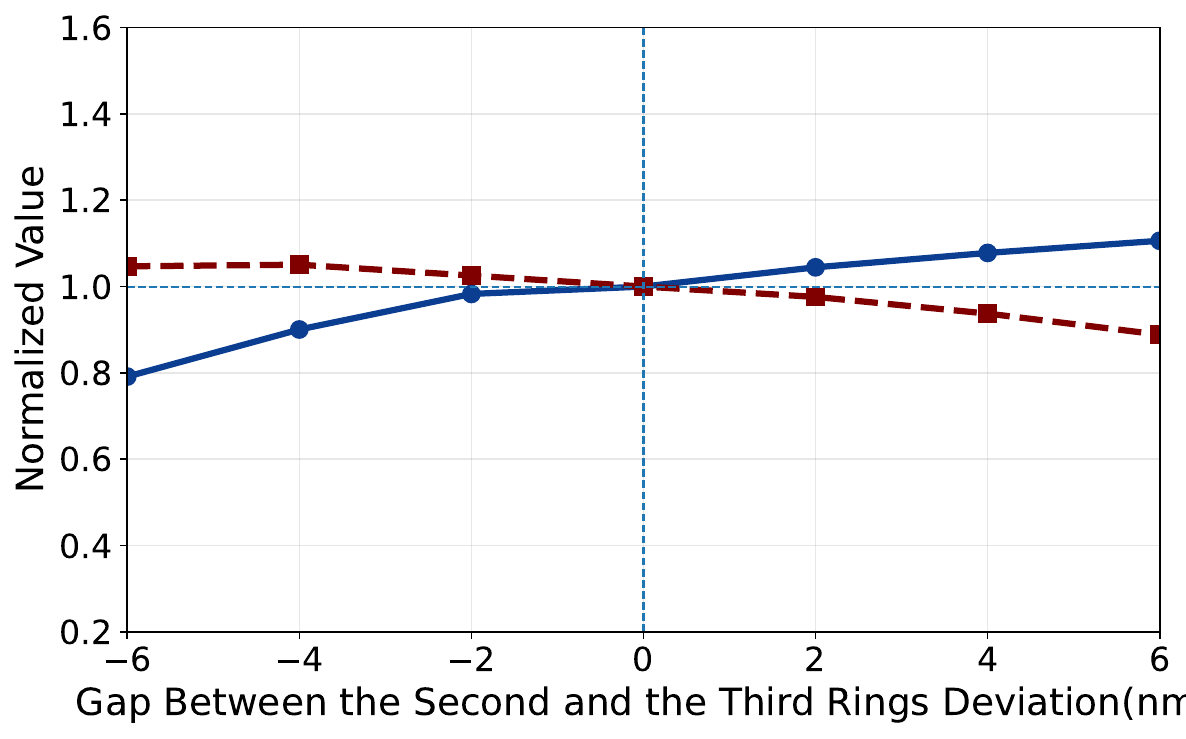}
        \caption{}
        \label{second-ring-gap}
    \end{subfigure}
    \caption{Single-parameter fabrication tolerance analysis of the inverse-designed cavity. Each panel shows the normalized Q-factor (solid blue line) and the collection efficiency  (dashed red line) as a function of the deviation of one geometric parameter from its optimized value: (a) First ring width deviation ($\Delta w_1$), (b) Second ring width deviation ($\Delta w_2$), (c) Third ring width deviation ($\Delta w_3$), (d) Central disk radius deviation ($\Delta r$), (e) Gap between the first and second rings deviation ($\Delta d_1$), (f) Gap between the second and third rings deviation ($\Delta d_2$).}
    \label{Fabrication-Tolerance}
\end{figure}

After obtaining an inverse-designed cavity with the desired figures of merit, we now examine the robustness of the cavity against dimensional variations introduced by potential nanofabrication errors. The nanofabrication process typically involves electron-beam lithography and dry etching that may introduce geometric uncertainties on the order of a few nanometers in feature dimensions. To study the sensitivity of the inverse-designed cavity to such variations, we perform a systematic single-parameter fabrication tolerance analysis. The analysis involves the FDTD simulation of the Q-factor and the collection efficiency while varying each geometric parameter of the structure over a range of $\pm 6$ nm from its optimized value, while keeping other parameters fixed at their optimized values. Figure \ref{Fabrication-Tolerance} presents such a tolerance analysis for the deviations of six geometric parameters compared to the ones of the inverse design: the widths of the first three rings ($\Delta w_1, \Delta w_2, \Delta w_3$), the central disk radius ($\Delta r$), and the first two inner ring gaps ($\Delta d_1, \Delta d_2$). The trends of all deviated geometrical features exhibit an inverse relation between the Q-factor and the collection efficiency. This inverse relation is expected because a higher Q-factor can be associated with lower losses, and consequently less emission to the far-field, while lower Q-factor can indicate stronger upward emission that translates to a greater collection efficiency. This trend is most noticeable for the first ring width, $\Delta w_1$ (Figure \ref{first-ring-width}): narrowing the ring by $6$ nm yields an enhancement of up to $40 \%$ in the Q-factor, together with $45 \%$ of the collection efficiency. Meanwhile, deviating other geometrical features across most of the range leads to a reduction of up to $20 \%$ in the Q-factor correlated with an increase of up to $20 \%$ collection efficiency and vice-versa. Such a reasonable trade-off with modest changes based on fabrication imperfection suggests that our inverse-designed cavity provides a good compromise between the Q-factor and collection efficiency that will not be substantially degraded due to small fabrication errors.

\section{Conclusions}

In this work, we demonstrate a gradient-based inverse design framework for using GME with automatic differentiation to co-optimize the Q-factor and the far-field emission pattern of a circular photonic cavity of rings that supports arbitrary polarization. Utilizing this framework, we obtain a novel circular cavity that exhibits almost an order-of-magnitude improvement in the Q-factor compared to the periodic bullseye cavity while preserving a nearly-Gaussian, directional far-field mode. Through a dimensional sensitivity analysis, we demonstrate that the designed cavity is reasonably robust to fabrication errors. While we considered the substrate material to be GaAs due to straightforward fabrication processes in this material as well as its ability to host InAs/GaAs quantum dots with efficient single photon emission capabilities, our approach can be adopted to other solid-state single photon emitting platforms such as color centers (e.g. in diamond, silicon, and SiC) \cite{SinglePhSiC, EnglundNVCenter} and rare earth ions. \cite{SinglePhEr} Furthermore, the GME approach for inverse design can be extended for other cavity geometries such as circular cavities with etched holes that was shown to offer high Q-factors without the use of bridges. \cite{Akahane2003, Yoshie2004, Li2021} The potential coupling of optically active matter qubits to high-Q cavities with efficient far-field emission modes obtained from or GME-based inverse design approach can pave the way toward establishing a high-cooperativity spin-photon interface toward the demonstration of deterministic photonic entanglement for quantum networking, distributed sensing, and measurement-based quantum computing. 






\vspace{10cm}
\printbibliography

\end{document}